\documentclass[twocolumn]{aastex631}
\usepackage{graphicx}
\usepackage{bm}
\usepackage{amssymb}
\usepackage{amsmath}
\usepackage{units}
\usepackage{array}
\usepackage{ulem}

\submitjournal{ApJ}
\received{December 4, 2023}
\revised{January 31, 2024}
\accepted{January 31, 2024}
\published{March 14, 2024}

\shorttitle{Disk induced binary precession}
\shortauthors{Tiede et al.}

\graphicspath{{./}{Figures/}}

\usepackage[T1]{fontenc}
\usepackage{aecompl}
\usepackage{calligra}
\DeclareMathAlphabet{\mathcalligra}{T1}{calligra}{m}{n}
\DeclareFontShape{T1}{calligra}{m}{n}{<->s*[2.2]callig15}{}

\usepackage{hyperref}
\hypersetup{
    pdfnewwindow=true,      
    colorlinks=true,       
    linkcolor=orange,      
    citecolor=cyan,        
    filecolor=orange,      
    urlcolor=orange           
}
\usepackage{units}
\usepackage{color}

\usepackage{color}

\def\Msun{{M_{\odot}}}

\def\GW{\mathrm{GW}}

\def\adot{\dot{a}}
\def\edot{\dot{e}}
\def\Mdot{\dot{M}}

\def\SNR{\textrm{SNR}}

\newcommand{\pomega}[1]{\dot \varpi_{\rm #1}}

\def\DD21{\citetalias{DOrazioDuffell:2021}}

\def\eg{\textit{e.g.}}

\defcitealias{DOrazioDuffell:2021}{DD21}

\begin{document}
\title{Disk induced binary precession: \\
Implications for dynamics and multi-messenger observations of black hole binaries}

\author[0000-0002-3820-2404]{Christopher Tiede}
\affiliation{Niels Bohr International Academy, Niels Bohr Institute, Blegdamsvej 17, 2100 Copenhagen, Denmark}

\author[0000-0002-1271-6247]{Daniel J. D'Orazio}
\affiliation{Niels Bohr International Academy, Niels Bohr Institute, Blegdamsvej 17, 2100 Copenhagen, Denmark}

\author[0000-0003-4818-3400]{Lorenz Zwick}
\affiliation{Niels Bohr International Academy, Niels Bohr Institute, Blegdamsvej 17, 2100 Copenhagen, Denmark}

\author[0000-0001-7626-9629]{Paul C. Duffell}
\affiliation{Department of Physics and Astronomy, Purdue University, 525 Northwestern Avenue, West Lafayette, IN 47907, USA}

%
\begin{abstract}
\noindent
Many studies have recently documented the orbital response of eccentric binaries accreting from thin circumbinary disks, characterizing the change in binary semi-major axis and eccentricity.
We extend these calculations to include the precession of the binary's longitude of periapse induced by the circumbinary disk, and we characterize this precession continuously with binary eccentricity $e_b$ for equal mass components.
This disk-induced apsidal precession is prograde with a weak dependence on binary eccentricity when $e_b \lesssim 0.4$ and decreases approximately linearly for $e_b \gtrsim 0.4$; yet at all $e_b$ binary precession is faster than the rates of change to the semi-major axis and eccentricity by an order of magnitude.
We estimate that such precession effects are likely most important for sub-parsec separated binaries with masses $\lesssim 10^7 \Msun$, like LISA precursors.
We find that accreting, equal-mass LISA binaries with $M < 10^6 \Msun$ (and the most massive $M \sim 10^7 \Msun$ binaries out to $z \sim 3$) may acquire a detectable phase offset due to the disk-induced precession.
Moreover, disk-induced precession can compete with General Relativistic precession in vacuum, making it important for observer-dependent electromagnetic searches for accreting massive binaries---like Doppler boost and binary self-lensing models---after potentially only a few orbital periods.
\end{abstract}
%

%
%

\section{Introduction} \label{S:Introduction}

Circumbinary accretion is astrophysically important for a variety of binaries ranging from protoplanetary systems, to binary stars, to massive black holes binaries.
In recent years, much numerical work has been performed to determine both the binary orbital and disk morphologic response to in-plane, prograde accretion \citep{MacFadyen:2008, Cuadra:2009, ShiKrolik:2012, ShiKrolik:2015, D'Orazio:CBDTrans:2016, MunozLai:2016, MirandaLai+2017, Yike+2017, Moody:2019, Tiede:2020, MunozLithwick:2020, Duffell:2020, HeathNixon:2020, Dittmann:2022:MachStudy, Franchini_SG:2021}.
Many of these works have focused primarily on equal-mass binaries with circular orbits, but the most recent studies have begun characterizing accreting binary systems across binary orbital eccentricities.
In particular, both \cite{Zrake+2021} and \cite{DOrazioDuffell:2021} found that while near-circular binaries with eccentricity $e \lesssim 0.1$ have their eccentricity damped towards orbital circularity--where the accretion flow causes them to expand their orbit--all other initial binary orbital eccentricities $e \gtrsim 0.1$ are driven towards an equilibrium eccentricity $e_{\rm eq} \sim 0.4 - 0.45$; and at $e_{\rm eq}$ the disk causes the binary to shrink its semi-major axis.
This is in agreement with three values of the eccentricity studied in \cite{Munoz:2019}.
\cite{Siwek:2023} established that this general phenomenon holds true for all binary mass ratios $q > 0.1$ with an equilibrium eccentricity that can vary from $0.25 \lesssim e_{\rm eq} \lesssim 0.5$ (but, however, did not find a circularizing regime at small $e$).
\cite{TiedeDorazio:2023} additionally investigated the orbital response of eccentric binaries in retrograde disks and found, contrary to prograde solutions, that the binary orbital eccentricity grows and the semi-major axis shrinks at all eccentricities $e \leq 0.8$.

For all of these studies, measurements of the binary orbital response to accretion from a circumbinary disk is focused on the disk mediated rate of change to the binary semi-major axis, orbital eccentricity, and mass ratio \citep[][]{Duffell:2020, Dittmann:2023:Q-Study}. 
However, in order to fully characterize disk driven alterations to the binary orbit, one must also consider binary apsidal precession induced by the gas forces. 
Such effects have recently been noted in simulations of stellar-mass binaries embedded inside AGN disks \citep{DittmannDempsey2023:smBHBs-AGN, Calcino+Ditt+Demp:2023:ECCsmMBHsAGN}, but have not been addressed in detail from full steady-state solutions of isolated binaries accreting from thin disks.
However, the presence of disk-induced binary precession has the potential to alter existing solutions for the orbital evolution of accreting binaries as well as to leave detectable effects in the observations of these systems with electromagnetic and gravitational waves (GWs).

Using existing data from \cite{DOrazioDuffell:2021}, in this paper we calculate the disk induced apsidal precession of the binary continuously for all eccentricities $0 < e < 0.9$.
In Section \ref{S:Methods} we briefly describe the simulations that underpin our analysis and lay out our framework for calculating and contextualizing the induced binary precession.
In Section \ref{S:Results} we present our computations for the disk induced binary precession and an analysis on its dynamical origin.
Section \ref{S:CBD-Modelling} details how and when such precession may be important for current and future modelling of accreting binary systems, and Section \ref{S:Obs-Implications} addresses how and when this precession might appear in observations of massive black hole binaries with periodic light curve searches and the space-based GW detector LISA.

\section{Methods} \label{S:Methods}

The data used in this study is the same as that from \citet[][hereafter \citetalias{DOrazioDuffell:2021}]{DOrazioDuffell:2021}. 
We highlight the most germane aspects of the system setup and numerical solution below and refer the reader to \citetalias{DOrazioDuffell:2021} and \cite{Duffell:2020} for more specific details.

Data was generated using the grid-based, moving-mesh hydrodynamics code \texttt{DISCO} to solve the 2D equations for viscous, locally isothermal hydrodynamics in the presence of a time varying binary potential. 
The binary has equal-mass components and is always fixed on a Keplerian orbit.
The circumbinary disk is treated in the thin-disk limit with aspect ratio $h/r \sim \mathcal{M}^{-1} = 0.1$ (where $\mathcal{M}$ is the Mach number of the flow).
The disk viscosity is chosen as a constant kinematic viscosity $\nu = 10^{-3} a^2 \Omega$ where $a$ is the binary semi-major axis and $\Omega$ is the binary orbital frequency.
The system is initially fixed on a circular orbit where it is run for 500 orbits to reach a quasi-steady configuration; and then the eccentricity of the binary $e_b$ is increased adiabatically up to $e_b=0.9$ over $2\times 10^4$ binary orbits.

As a diagnostic, \texttt{DISCO} outputs all forces $\mathbf{f}$ on the binary due to the gas.
The precession of the binary's longitude of periapse can be calculated from these forces \citep{Murray:1994} as 
\begin{align}
    \dot \varpi_b =  \sqrt{ \frac{a}{GM} \frac{(1 - e_b^2)}{e_b^2}} \left[ -f_r \cos{\nu} + f_\phi \sin{\nu} \frac{2 + e_b \cos{\nu}}{1 + e_b \cos{\nu}} \right] \ .
 \label{eq:pomdot}
\end{align}
where $M$ is the total binary mass and $\nu$ is the orbital true anomaly.
There is no contribution to this precession from the accretion of mass ($\dot M$) and we demonstrate this in Appendix \ref{A:derivation}. There is a small contribution from the direct accretion of momentum, but these effects are subdominant to the gravitational forces from the circumbinary material; so all effects herein are calculated only from the gravitational forces on the binary.

In order to translate this precession into meaningful units, we re-write Equation~\ref{eq:pomdot} as 
\begin{align}
 \begin{split}
    \pomega{b} &= \frac{G\Sigma_0}{v_b} \sqrt{\frac{1-e_b^2}{e_b^2}} \left( \tilde f_{\phi}\sin{\nu} \frac{2+e_b\cos{\nu}}{1+e_b\cos{\nu}} - \tilde f_r \cos{\nu} \right)  \\
    &= \frac{G\Sigma_0}{v_b} \pomega{sim}
 \end{split}
\end{align}
where $\Sigma_0$ is the density scale and $v_b = \sqrt{GM / a}$ is the average binary orbital velocity. $\tilde f_i$ denotes the specific forces (accelerations) measured in code-units where $GM = \Sigma_0 = a = 1$. $\pomega{sim}$ is the measured precession rate in code units.
Therefore, the physical binary precession rate from the circumbinary disk can be written in terms of the binary orbital frequency $\Omega_b = \sqrt{GM / a^3}$ as
\begin{align}
    \pomega{b} = \frac{\Sigma_0 a^2}{M} \pomega{sim} \Omega_b = q_D \pomega{sim} \Omega_b
    \label{eq:pomega-b}
\end{align}
where $q_D = \Sigma_0 a^2 / M$ is the disk-to-binary mass ratio. 
The magnitude of the binary precession relative to other relevant timescales (discussed below) depends on the magnitude of $q_D$.
Because the underlying simulations are scale free (i.e. are true for all mass and length scales $a$, $M$, and $\Sigma_0$)\footnote{To the extent that the assumptions of a thin, radiatively efficient, and gravitationally stable disk hold true; this is discussed further in Section \ref{S:CBD-Modelling}.}, derivatives of the binary orbital elements are typically reported per unit mass accreted (in units of $\dot M$) as opposed to per unit time.
Thus, we can also express the precession rate per accreted mass by noting that in steady-state $\Sigma_0 = \dot M_0 / 3 \pi \nu$ with $\dot M_0$ the accretion rate through the disk and $\nu = \nu_0 a^2 \Omega_b$ such that
\begin{align}
    \pomega{b}  = \frac{\pomega{sim}}{3\pi\nu_0} \frac{\dot M_0}{M}  \quad\text{and}\quad q_D =   \frac{1}{3\pi\nu_0} \frac{\dot M_0}{M} \,\Omega_b^{-1}  \ .
 \label{eq:pomega-dm}
\end{align}
%

\section{Results} \label{S:Results}

\begin{figure}
    \centering
    \includegraphics{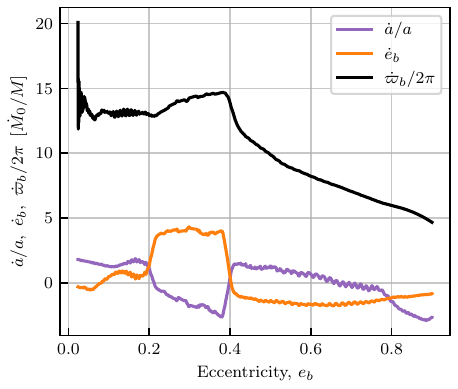}
    \vspace{-5pt}
    \caption{Change in binary orbital elements per unit accreted mass: longitude of periapse (\emph{black}), semi-major axis (\emph{purple}), and eccentricity (\emph{orange}). We see that binary precession occurs on an order of magnitude faster timescale than changes to the binary semi-major axis or eccentricity.
    We note that these values can be mapped to units that depend on the disk mass $q_D$ and binary orbital frequency through Equation~\ref{eq:pomega-dm}.
    }
 \label{fig:orb-evo}
\end{figure}

The primary finding of \DD21 was the continuous characterization of how an accreting binary changes its semi-major axis $\dot a / a$ and eccentricity $\dot e_b$ as a function of binary eccentricity.
In Figure~\ref{fig:orb-evo} we plot these same curves for $\dot a / a$ and $\edot_b$ and add the new measurement of the binary apsidal precession rate $\pomega{b} / 2\pi$ according to Equation~\ref{eq:pomega-dm}.
Apsidal precession is shown by the \emph{black} curve, the change in semi-major axis by the \emph{purple}, and the change to eccentricity in \emph{orange}.
We observe that disk-induced binary precession is \emph{prograde} and approximately constant for $e_b \lesssim 0.3$, peaks at $e_b \approx 0.4$, and decays approximately linearly for $e_b \gtrsim 0.4$. 
In Appendix \ref{A:retro} we also show the binary apsidal precession induced by a \emph{retrograde} CBD \citep{TiedeDorazio:2023} noting that such scenarios only change the precession rate by at most a factor of two\footnote{One might also consider inclined disks where the interplay of binary and disk eccentricity with binary and disk inclination can follow more complex, Kozai-Lidov-type oscillations that are tied with the disk-induced precession \citep[c.f.][]{Martin2023:MisalignedCBDs}.}.
We include approximate fitting functions for these curves in Appendix \ref{A:fit}.

\begin{figure}
    \centering
    \includegraphics{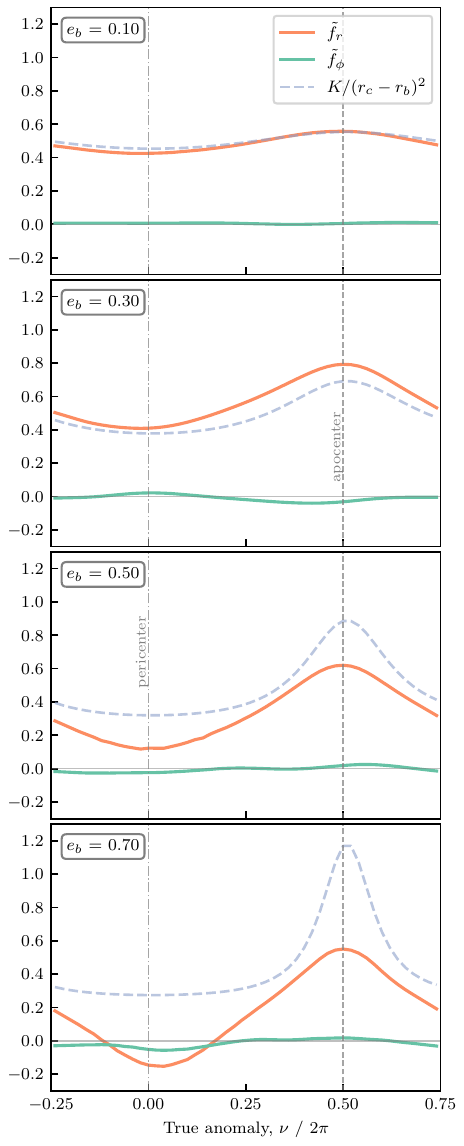}    
    \vspace{5pt}
    \caption{
    Average forces in the radial (\emph{orange}) and azimuthal (\emph{green}) directions at each binary phase (true anomaly, $\nu$).
    At nearly all phases, there is a large outward radial force from the bulk of the CBD which is the dominant contribution to the total force (except for at pericenter when $e_b$ is large; e.g. $e_b=0.7$).
    The \emph{dashed-blue} line shows the linear force in a toy model between 
    a point mass at the disk cavity radius $r_c = 3\, a$
    and a particle on a Keplerian orbit with eccentricity $e_b$ and radius $r_b = (1 - e_b^2) / (1 + e_b\cos{\nu})$ as it sweeps through true anomalies $\nu \in [0, 2\pi]$.
    }
 \label{fig:radial-forces}
\end{figure}

Of primary note, we see that the apsidal precession of the binary is a full order of magnitude faster than corresponding changes to the binary eccentricity and semi-major axis.
We can cultivate an intuitive understanding for this by more closely examining the forces on the binary.
Specifically, the disk always exerts a comparatively large outward radial force on the binary (except for short times at pericenter when $e_b$ is large).
This is evident in Figure~\ref{fig:radial-forces} which shows the average radial (\emph{orange} curve) and azimuthal (\emph{green} curve) force at each binary phase (here given by the orbital true anomaly $\nu$) for four representative eccentricities $e_b = \{0.1, 0.3, 0.5, 0.7\}$.
Generally speaking, the radial force is largest near binary apocenter and smallest near pericenter because these are when the components are nearest and furthest, respectively, from the CBD cavity edge and the bulk of the disk.
To illustrate this, a generalized gravitational force $f = K / (r_c - r_b)^2$ between a point mass placed at the disk cavity radius $r_c = 3a$ and another particle on a Keplerian orbit with phase-dependent radius $r_b = a(1 - e^2) / (1 + e \cos{\nu})$ with the constant arbitrarily chosen as $K = 2$.
However, at larger eccentricities, this approximation breaks down because of increasing amounts of persistent intra-orbit material.

\begin{figure}
    \centering
    \includegraphics{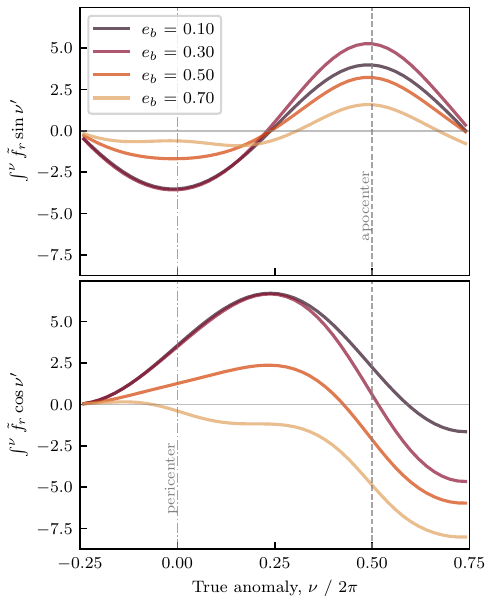}
    \vspace{-5pt}
    \caption{
    The cumulative sum of the radial force times the sine (\emph{top}) and cosine (\emph{bottom}) of the binary phase. 
    The evolution of the binary semi-major axis and eccentricity are proportional to the former, and we see that the sum over a full binary orbit is comparatively very small because of the predominantly symmetric nature of the radial force.
    The precession of the binary's longitude of pericenter, however, is proportional to the latter, which remains comparatively large when summed over a full orbit.
    }
 \label{fig:integrated-forces}
\end{figure}

\begin{figure*}
    \centering
    \includegraphics{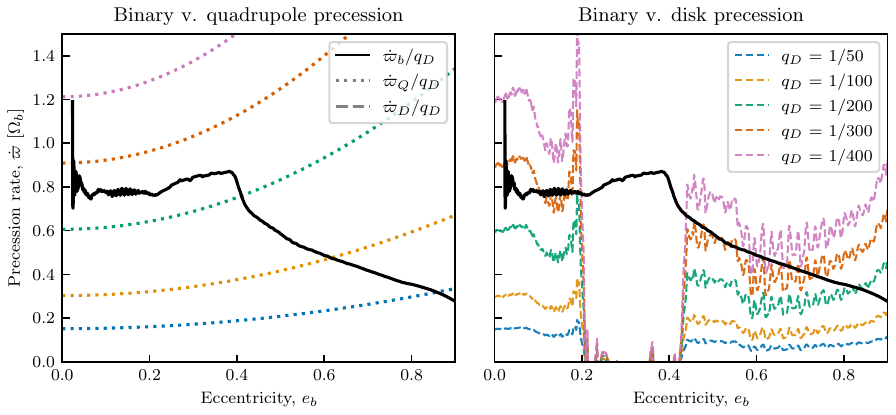}
    \vspace{-5pt}
    \caption{The binary precession rate due to gravitational forces from the circumbinary disk $\pomega{b}$ compared with the quadrupolar precession of the disk itself $\pomega{Q}$ (\emph{left} panel; \emph{dotted, colored} curves) as well as the empirically measured precession of the disk $\pomega{D}$ (\emph{right} panel; \emph{dashed, colored} curves) at different disk-to-binary mass ratios $q_D$.
    In both scenarios, $\pomega{b}$ competes with the precession of the disk when $q_D \sim \mathcal{O}(10^{-2})$; and disk-induced precession becomes the dominant effect at all eccentricities when $q_D \gtrsim 1 / 50$.
    }
 \label{fig:binary-precession}
\end{figure*}

The large radial force does not contribute significantly to $\adot$ and $\edot_b$ because these are only affected by torques ($T = r f_{\phi}$) and components of the work $W \sim f_r v_r \propto f_r \sin{\nu}$ (see Appendix A).
The latter is almost always small because $v_r$ is anti-symmetric around pericenter/apocenter ($\sin \nu$ dependence), whereas the large outward portion of $f_r$ due to the bulk of the disk is symmetric around the same points (with varying $r(\nu)$). 
Thus, only a small anti-symmetric part of $f_r$ contributes to $\adot$ and $\edot$.
This can be seen in the top panel of Figure~\ref{fig:integrated-forces} which shows the cumulative sum
\begin{align*}
    \int_{-\pi/2}^\nu \, \tilde f_r \, \sin{\nu^\prime} \,d\nu^\prime 
\end{align*}
(in shorthand, $\int^\nu \tilde f_r \sin{\nu^\prime}$).
One can see that the orbit average of $f_r \sin\nu$ is nearly zero, such that the large radial force from the bulk of the CBD plays a subdominant role in $\dot a$ and $\edot_b$.
However, the precession rate is proportional to the binary acceleration as opposed to the orbital velocity, $\pomega{b} \propto f_r \cos{\nu}$ (see Equation~\ref{eq:pomega-b} and Appendix~\ref{A:derivation}).
The bottom panel of Figure~\ref{fig:integrated-forces} illustrates the corresponding orbital average $\int^\nu \tilde f_r \cos\nu^\prime$ which is notably larger than the purely anti-symmetric component giving a substantial contribution to the binary precession.
Therefore, as the binary moves from the point of maximal outward radial velocity, through apocenter, to the point of maximal negative inward velocity, it is decelerating, and the outward radial force fights this; hence delaying the apocenter turning point.
This slows the radial oscillations of the orbit compared to the azimuthal oscillations advancing the pericenter angle. 
The opposite happens during the accelerating portion of the binary orbit.
The force near apocenter is stronger than that at pericenter because the binary is closer to the circumbinary disk, and this mismatch leads to pericenter advance prograde with the orbit. 
Interestingly the $f_r \cos{\nu}$ symmetric part of $f_r$ does no work on the binary, and rather---in analogy to the magnetic portion of the electromagnetic field tensor---only changes its orientation via $\pomega{b}$.

The trend towards constant time averaged $\pomega{b}$ at small $e$ can be understood by expanding the dominant $f_r$ term in Equation~\ref{eq:pomdot} in terms of the mean anomaly $\tilde M(t)$,
\begin{equation*}
    -   \frac{f_r}{e} \cos{\nu} = -\frac{f_r \cos{\tilde M}}{e}  - f_r \left(\cos{2\tilde M} - 1\right) - \mathcal{O}(e).
\end{equation*}
Averaging the above over an orbit will yield the time averaged value of the precession rate $\left< \pomega{b} \right>_P$.
As seen in the top panel of Figure~\ref{fig:radial-forces}, at small $e_b$ we can approximate $f_r = A_r(1-e\cos{\tilde M})$ with $A_r$ constant. 
The orbital average, then, is
\begin{equation}
    \left< \pomega{b} \right>_P \approx \sqrt{ \frac{a}{GM} } \left(\frac{3}{2}A_r \right) \approx 0.75 q_D \Omega_b,
\end{equation}
which for $A_r\sim 0.5$, from the top panel of Figure~\ref{fig:radial-forces}, corresponds to the value plotted in Figure~\ref{fig:orb-evo} at $e\lesssim0.2$ (by Equation~\ref{eq:pomega-dm}, $13 / 2\pi \times \Mdot_0 / M \approx 0.77\, q_D \Omega_b$; see Figure~\ref{fig:binary-precession}).

Lastly, we note that because the precession rate is dominated by the outward radial pull of the CBD, we expect our results to be comparatively insensitive to system uncertainties like the disk thermodynamics (unlike the other orbital elements; see e.g. \citealt{Tiede:2020, Dittmann:2022:MachStudy, WangBaiLai:2023}) or whether the binary is prograde or retrograde (Appendix~\ref{A:retro}).

\section{Implications for CBD modelling} \label{S:CBD-Modelling}
%

\subsection{Binary vs. disk precession} \label{s:binary-v-disk}
%

\begin{figure}
    \centering
    \includegraphics{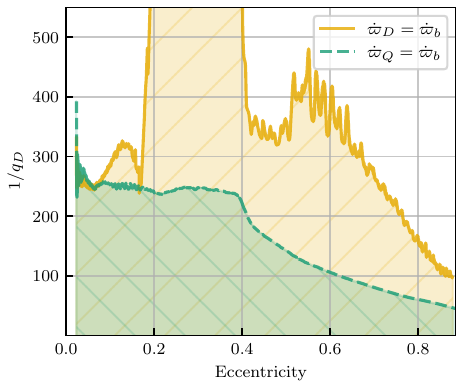}
    \vspace{-5pt}
    \caption{The critical value of the binary-to-disk mass ratio $1/q_D$ where the binary precession rate $\pomega{b}$ is equal to the quadrupolar precession rate (\emph{green, dashed)} at the disk cavity edge $r_c=3.25\,a$ and the measured disk precession (\emph{gold, solid)}.
    Below each curve $\pomega{b}$ dominates.
    Between $0.18 \lesssim e_b \lesssim 0.4$, the critical binary-to-disk mass ratio $1 / q_D$ goes to infinity because the disk no longer precesses.
    In this regime, the bounding limit is a viscous time at the disk's cavity edge which yields a critical value $1 / q_D \approx 4.8 \times 10^3$. 
    }
    \label{fig:qd-crit-pom-disk}
\end{figure}

Most prior solutions for eccentric binaries accreting from steady-state disks assume that the binary is on a fixed Keplerian orbit and does not precess.\footnote{Although solutions from Smoothed Particle Hydrodynamics studies that directly integrate the binary orbit ought to observe this effect for large enough disk masses.} 
A primary finding for circular (and low-$e_b$) systems is that the CBD becomes eccentric and precesses around the binary with a frequency comparable to that induced by the binary potential's quadrupole moment \citep[e.g.][]{MacFadyen:2008, ShiKrolik:2012, MunozLithwick:2020}
\begin{equation}
    \pomega{Q} \approx \frac{3}{4}\frac{q}{(1 + q)^2} \left( 1 + \frac{3}{2}e_b^2 \right) \left( \frac{r_c}{a} \right)^{-7/2} \Omega_b \ 
\end{equation}
where $q$ is the binary mass-ratio and $r_c$ is the approximate radius of the disk cavity.
\DD21 found that this holds for eccentric binaries $e_b \lesssim 0.2$, but that for $0.2 \lesssim e_b \lesssim 0.4$ the disk undergoes a state-transition to a predominantly symmetric, non-precessing configuration; and for $e_b \gtrsim 0.5$ the disk becomes highly eccentric and precesses around the binary slower than the associated frequency from the binary quadrupole moment.
However, it is conceivable that such behavior could be altered if the binary is also precessing at a comparable rate.

In order to determine the relative importance of binary precession compared to the disk precession itself, we compare the binary precession rate to that of a test particle in the quadrupole potential of the binary and to the measured disk precession rate $\pomega{D}$ from \DD21.
This is illustrated in Figure~\ref{fig:binary-precession} where the black curve is equivalent to that in Figure~\ref{fig:orb-evo}, except given in units according to Equation~\ref{eq:pomega-b}.
The colored curves in the left panel show the precession rate from the binary quadrupole moment with $q = 1$ and $r_c = 3.25a$ scaled into equivalent units through $q_D$.
The curves in the right panel show the same comparison to the measured disk precession rate, again scaled to equivalent units through $q_D$.
The displayed values of $q_D$ were chosen to highlight values of the disk-to-binary mass ratio where the disk-induced binary precession might compete with the precession of the disk itself.
In particular we see that for disk masses $M_D \lesssim M / 400$ the induced binary precession is always subdominant to the precession of the disk itself (and quadrupolar precession at $r_c = 3.25a$).
However, for disk masses $M_D \gtrsim M / 400$ the induced binary precession can compete with the precession of the disk at most eccentricities, and for masses $M_D \gtrsim M / 100$ the binary precession is always faster than that of the disk.
In these regimes, the effects of binary precession may be significant for the full hydrodynamics solution including the disk precession itself.
The exceptions to this are for $0.2 \lesssim e_b \lesssim 0.4$ when the disk ceases to precess, causing the binary precession to always dominate in this regime.

For intermediate values of the disk mass $1/400 \lesssim q_D \lesssim 1/50$ we can solve for the critical value where the binary precession rate equals that of the quadrupolar rate and disk rate respectively; these values are shown as a function of eccentricity in Figure~\ref{fig:qd-crit-pom-disk}.
The \emph{green-dashed} curve shows the critical values of $1 / q_D$ for quadrupolar precession at $r_c = 3.25a$, and the \emph{gold-solid} curve illustrates the critical values for the empirical disk precession.
The shaded regions below each curve show disk masses where binary precession dominates.
We see, again, that because the disk ceases precession for $0.2 \lesssim e_b \lesssim 0.4$ the critical value of $1 / q_D$ diverges and binary precession dominates always.
In this case, the bounding timescale would be the viscous time in the disk; i.e. if binary precession is fast and the orientation of the gravitational potential rotates faster than the disk can relax in response to their changing relative orientation.

\subsection{The viscous limit} \label{s:viscous-limit}

\begin{figure}
    \centering
    \includegraphics{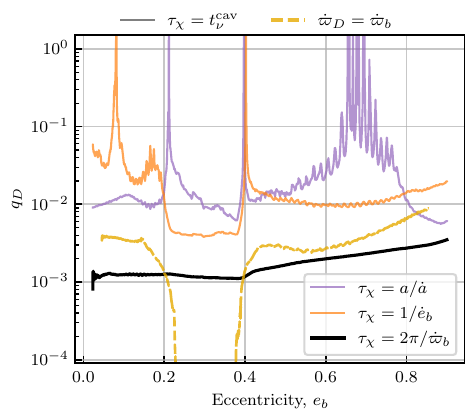}
    \vspace{-5pt}
    \caption{Disk-to-binary mass ratios where $\tau_\chi$ equals the viscous time at the cavity edge $t_\nu^{\rm cav}$ as a function of binary eccentricity. The \emph{dashed-gold} line shows the critical value of $q_D$ where disk-induced binary precession rate equals the precession rate of the eccentric disk $\pomega{b} = \pomega{D}$.
    }
 \label{fig:qd-visc}
\end{figure}

Similar to Section \ref{s:binary-v-disk}, we can compute an estimate of the disk mass where binary precession (and the other evolution effects) may influence the full solution by equating their characteristic timescales to the viscous time in the disk.
That is to say, we compute the disk mass where the binary changes its orbit elements faster than the disk can relax and communicate these changes viscously.
We compute the characteristic number of orbits $\tau_\chi$ required for an order-unity change to each of the binary orbital elements $\chi = \{\dot a / a, \edot_b, \pomega{b} / 2\pi \}$ through Equation~\ref{eq:pomega-b} (and as the inverse of the curves shown in Figure~\ref{fig:orb-evo}).
We note that we can generalize Equation~\ref{eq:pomega-b} to $\dot \chi = q_d \dot \chi_{\rm sim} \Omega_b$, such that the timescale associated with each $\dot \chi$ is set by $q_D$.
Thus, in Figure~\ref{fig:qd-visc} we compute the value of $q_D$ where the timescale for changing each orbital element equals the viscous time at the cavity wall $\tau_\chi = t_\nu^{\rm cav}$ as a function of the binary eccentricity.
We see that the limit for the disk's ability to relax viscously to changes in $\dot a / a$ and $\edot_b$ occurs at disk masses $M_D \approx 10^{-2} M$; but that the binary precession begins out-running the viscous time for order-of-magnitude less massive disks, $M_D \approx 10^{-3} M$.
Therefore, for disk masses $M_D \gtrsim 10^{-3} M$ it may be important to include the disk-induced binary precession (i.e. by integrating the binary orbit in response to forces from the CBD) in order to fully characterize the solution.
For completeness, we also overlay the limit where the binary precession equals the empirical disk precession from Section \ref{s:binary-v-disk} as the \emph{blue-dotted} curve.

\begin{figure*}
    \centering
    \includegraphics{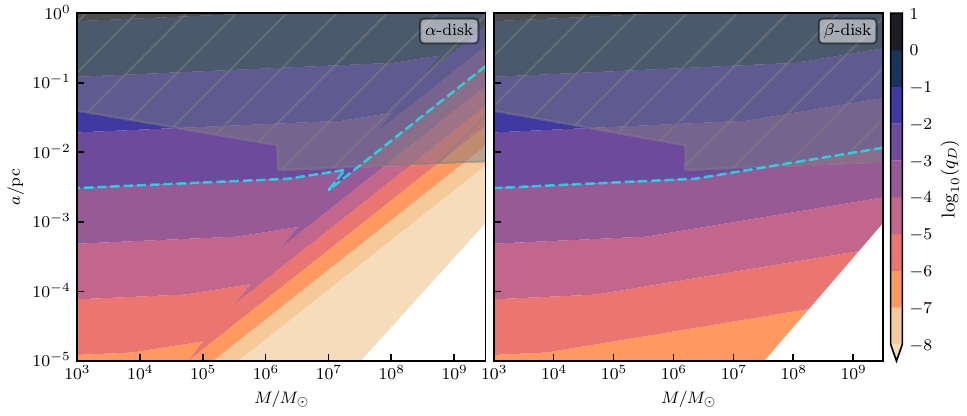}
    \vspace{-5pt}
    \caption{Contours of $q_D = \Sigma_k a^2 / M$ from steady-state solutions for $\alpha$- and $\beta$-disks (\emph{left} and \emph{right} respectively, \citealt[][eqs. 4-14]{Haiman+2009}) accreting at their Eddington limit with $\Sigma_k = \Sigma_{(\alpha,\,\beta)}(a)$.
    The \emph{dashed-cyan} lines shows $q_D = 10^{-3}$ near the "viscous limit" determined in Section \ref{s:viscous-limit}.
    Above this line, precession effects may alter the full solution.
    The grey shaded region in the upper portion of the plot 
    illustrates the limit where disks are no-longer stable against their own gravity.
    The white region in the lower right is where the semi-major axis is within the ISCO of the binary components.
    We see that disk-induced binary precession from a gravitationally stable disk could be important for the dynamics of binaries with $M \lesssim 10^7 \Msun$, which are likely progenitors of mergers in the LISA band. 
    }
  \label{fig:ab_disk_qd}
\end{figure*}

\subsection{Disk mass from steady-state models} \label{s:steady-state-disks}
Lastly, steady-state solutions for thin accretion disks around single black holes allow one to directly calculate the mass of the surrounding disk.
Thus, using Equations 4-14 from \cite{Haiman+2009} we calculate the disk-to-binary mass ratio $q_D = \Sigma_k a^2 / M$ for two steady-state solutions: \emph{$\alpha$-disk} where the disk viscosity is proportional to the total pressure (gas + radiation), and \emph{$\beta$-disk} where the viscosity is only proportional to the gas pressure.
Such solutions consider a disk with three distinct regions: an inner-region that is radiation pressure dominated with opacity given by electron-scattering, a middle-region that is gas-pressure dominated with electron-scattering opacity, and an outer-region that is gas-pressure dominated with free-free opacity.
The corresponding disk density is then taken as $\Sigma_k = \Sigma_{(\alpha ,\, \beta)}(r=a)$.
Figure~\ref{fig:ab_disk_qd} shows values of the disk-to-binary mass ratio $q_D$ for binary's accreting at the Eddington rate with mass $M/\Msun \in [10^3 - 10^{10}]$ and separations $a \in [10^{-5}, 1]$ parsec (where we've chosen the disk viscosity according to the $\alpha$-prescription with $\alpha = 0.3$ \citealt{SS73}).
The \emph{left} panel shows solutions for $\alpha$-disks and the \emph{right} for $\beta$-disks.
The \emph{hatched, grey} region illustrates where such a steady-state solution is no longer gravitationally stable and the \emph{white} region in the bottom right corner shows where the ISCO of the binary components is larger than the semi-major axis.
The \emph{dashed, cyan} line shows the ``viscous limit'' $q_D = 10^{-3}$ where binary precession may outpace the ability of the circumbinary disk to relax to the changing binary apsides.
Therefore, for binaries of mass up-to $10^7 M_\odot$ (likely LISA progenitors) that experience an active, gas accretion phase during their evolution through the sub-parsec regime, the full description of the binary evolution and associated observational signatures may require a solution that includes such precession effects.

\section{Observational implications} \label{S:Obs-Implications}

Binary precession can leave detectable imprints in both electromagnetic and gravitational wave emission from accreting or inspiraling binaries, especially in so far as it competes with precession from general relativity.
We explore these effects by comparing disk-induced precession rates to those from general relativity in vacuum.
The lines in Figure~\ref{fig:binary-v-gr} show the combinations of binary mass $M$ and semi-major axis $a$ where the rate of disk-induced precession is equal to the orbital precession from GR in vacuum $\pomega{b} = \pomega{GR}$, with
\begin{equation}
    \pomega{GR} = \frac{3GM}{a(1 - e_b^2)c^2} \Omega_b \ ,
\end{equation}
for varying binary eccentricity and a disk mass $q_D = 10^{-3}$.
The shaded regions show where binary precession dominates, and the white regions where GR precession is fastest.
The ratio of binary-to-GR precession can be written as
\begin{equation}
    \frac{\pomega{b}}{\pomega{GR}} = \frac{a(1 - e_b^2) c^2 q_D}{3GM} \pomega{sim}, \ 
\end{equation}
such that the colored curves in Figure~\ref{fig:binary-v-gr} simply depict the linear relationship between mass and semi-major axis at fixed $e_b$ when this ratio $\pomega{b} / \pomega{GR} = 1$; and the eccentricity dependence is set by the relationship in Figure~\ref{fig:orb-evo}.
The dashed black line shows the binary-disk decoupling radius where the viscous inflow rate at the disk's inner edge $v^{(\nu)}_{r}$ equals the orbital decay rate due to gravitational wave radiation $\dot a_{\rm GW}$ \citep[c.f.][]{ArmNat:2002:ApJL}.
For $q_D = 10^{-3}$, at the limit where the disk has sufficient time to viscously relax to the changing binary orbit,
GR precession becomes the dominant source of binary precession markedly before the binary decouples from the circumbinary disk.
For smaller values of $q_D$ this transition recedes to larger binary semi-major axes, while for larger $q_D$---although in this regime our solutions may no longer be applicable (see Section \ref{s:viscous-limit}
)---disk-induced precession encroaches on the decoupling radius; and it dominates over GR precession all the way down to the decoupling radius when $q_D \sim 0.1$.

The primary observable effect of apsidal precession, whether or not it dominates over other forms of precession, is the accumulation of extra orbital phase relative to an unperturbed orbit. 
We show below that this can arise in observations of periodic lightcurves from eccentric accreting binaries or in the dephasing of GW emission from supermassive black hole binaries. 
To lowest order in orbital eccentricity, the magnitude of this change over a period is
\begin{equation}
    \left<\delta \theta\right>_P \approx \pomega{b} P_b  \sim 2\pi q_D
    \label{eq:dephase-obs}
\end{equation}
where we note that $\pomega{b}$ and this dephasing do not necessarily vanish for very small or zero eccentricities because external forces can still alter the effective gravitational potential and the orbital velocity of the binary (see Appendix \ref{A:dphi}).

\subsection{Periodic Light Curve Searches for Black Hole Binaries}

\begin{figure}
    \centering
    \includegraphics{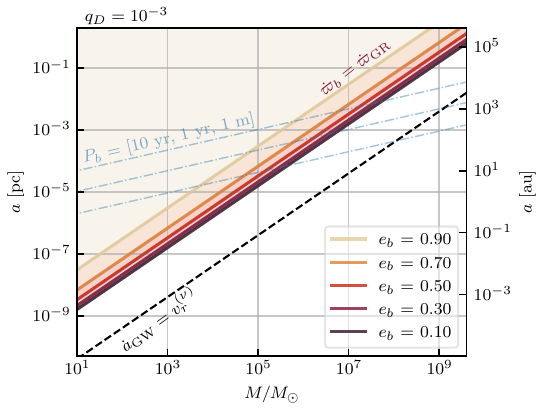}
    \vspace{-8pt}
    \caption{Contours of binary mass and separation where $\pomega{b} / \pomega{GR} = 1$ for 5 values of the binary eccentricity with a disk mass $q_D = 10^{-3}$.
    The lightly-shaded regions above the curves indicate where disk-mediated binary precession would dominate over GR-precession for each eccentricity.
    The black dashed line shows binary separations where orbital decay from the emission of GWs causes the binary to decouple from the CBD (e.g. is faster than the viscous inflow of the disk at its inner-edge, $r_c = 3.25a$).
    The \emph{blue, dash-dotted} lines show binary periods $P_b$ of 10 years, 1 year, and 1 month from top-to-bottom.
    }
  \label{fig:binary-v-gr}
\end{figure}

The relative contributions of disk-induced precession vs. general relativistic precession can be significant for electromagnetic searches for massive binaries in galactic nuclei.
A subset of search methods aim to identify periodic features in target system lightcurves that can be connected to the underlying orbital period of an accreting binary \citep[\eg,][]{Graham+2015a, Charisi+2016, LiuGezari+2019, Chen_DES_PLCs+2020, Chen_ZTF_PLCs+2022}.
Orbital periods targeted in these searches range from weeks to years \citep[\eg,][]{HKM09, XinHaiman_LSSTshort:2021, RomanWP:MBHBs}. For comparison, Figure~\ref{fig:binary-v-gr} shows \emph{blue, dashed-dotted} lines for three binaries periods, from top-to-bottom, $P_b \in$ [10 years, 1 year, 1 month]; like those that might be identifiable in electromagnetic searches.

Binary models for the origin of periodic variability employ either hydrodynamic variability caused by variations in the binary accretion rate \citep[e.g.][]{DHM:2013:MNRAS, Farris:2014, Yike:2018, Dittmann:2022:MachStudy, Westernacher-Schneider:2022, Gutirrez+2022} or observer-dependent relativistic Doppler Boosts \citep{PG1302Nature:2015b, DHLighthouse:2017, Charisi+2018} and binary self-lensing events \citep{DOrazio:2018:Lensing, Spikey:2020, KelleyLens+2021, Davelaar:2022:LensingFlares, 2023arXiv231019766M}---both modulated by the binary orbit.
Hydrodynamic accretion-rate variability would likely not be sensitive to disk-induced binary precession for $q_D \lesssim 10^{-3}$ when $\pomega{b}$ is slow compared to the other relevant timescales, but it is possible that the characteristic frequencies of accretion variability change in the limit $q_D \gtrsim 10^{-3}$ (i.e., Figure~\ref{fig:qd-visc}); we leave exploration of this effect to future work.

Relativistic boosting and lensing, however, are sensitive to---and straight-forwardly depend on---the orbital reconstruction. 
Specifically, the shape of periodic modulations caused by the orbital Doppler boost along with the shape, magnification, and crucially, timing of lensing flares depend on the orbital eccentricity and argument of pericenter with respect to the observer's line of sight \citep[see Fig. 7 in][]{DOrazioCharisi_Rev:2023}; and thus, also on changes to these quantities over time, such as the advance of pericenter.
These models sometimes include general relativistic precession in their modelling, but Figure~\ref{fig:binary-v-gr} demonstrates that a circumbinary disk sourcing the binary accretion with $q_D \sim 10^{-3}$ may induce a comparable effect for massive binaries $10^{4} \lesssim M / M_\odot \lesssim 10^{7}$ with orbital periods of order months-to-years.

To estimate when precession from a circumbinary disk would significantly affect the shape and timing of Doppler+lensing signatures, we calculate the number of binary orbits needed for the accumulated precession angle to equal the width of a binary self-lensing flare\footnote{This assumes at least one lensing flare has been temporally resolved and so deviation from the non-precessing model can be verified by measuring a later flare that is offset by greater than one flare width.}. 
This limit guarantees at least an observable change in flare timing, if not also a change in lightcurve shape due to an altered azimuthal viewing angle caused by apsidal precession.
From Equation~\ref{eq:dephase-obs} we equate the accumulated precession angle $\delta \theta \approx  2 \pi q_D N_{\rm orb}$ with the duration of the lensing event divided by the orbital period (\eg, Equation~(3) of \cite{DOrazio:2018:Lensing}).
Then the required number of orbits needed to accumulate significant precession is,
\begin{eqnarray}
        N_{\rm orb} &\gtrsim& \frac{(1+q_b)^{-1/2}}{\pi^{5/3} q_D } \left( \frac{2 GM}{c^3 P_{\rm b}}\right)^{1/3} \\ \nonumber
        &\approx& 1.2  \left( \frac{q_D}{10^{-3} } \right)^{-1} \left( \frac{M}{10^7 \Msun} \right)^{1/3} \left( \frac{P_{\rm b}}{2 \mathrm{yr}} \right)^{-1/3}.
\end{eqnarray}
In the second line we evaluate for $q_b=1$ and a binary at the edge of the disk-precession-dominated regime in Figure~\ref{fig:binary-v-gr}. 
Hence, before relativistic apsidal precession becomes dominant, the disk-induced precession can cause significant changes to the Doppler-lensing periodic modulations over only a few orbital periods. 
This motivates including disk orbital precession when modelling periodic variability due to binary self-lensing. 
Moreover, such lightcurve models with the inclusion of disk precession allow measurement of the local disk mass and can provide otherwise lacking constraints.

Forces from the disk which cause pericenter precession also alter the orbital velocity of the binary away from the Keplerian value.
In our case, the dominant contribution is from outward radial disk forces which do not strongly affect orbital evolution (see Section \ref{S:Results}), but do act to slow down the binary compared to the Keplerian value for a given separation and binary mass. 
For Doppler+lensing signatures this effect would bias the binary parameters that one would recover assuming an orbit in vacuum by, an albeit, very small amount of order $q_D$.

\subsection{Dephasing of Gravitational Waves in LISA}

Disk-induced precession in the late stages of massive binary evolution could also potentially have important consequences for near-equal-mass binary inspiral events that will be detected by LISA.
Binaries in the LISA band ($10^3 \lesssim M / M_\odot \lesssim 10^7$) can stay coupled to their circumbinary disks anywhere from $\sim 1$ year down to minutes prior to merger \citep{Dittmann:2023:Decoupling, Krauth:2023}; or they may not decouple at all \citep{Avara:3DMHD:2023}.
Therefore, even though the disk-induced precession in-band is likely small compared to that from GR,
the responsible disk-forces could still feasibly alter gravitational wave emission and could leave a detectable imprint on LISA signals.

In accordance with LISA expectations, we focus on sources with very low eccentricity\footnote{This is consistent with estimates for the in-band eccentricity $e_b \sim 10^{-3} - 10^{-4}$ for equal-mass binaries that have had their eccentricity driven near the equilibrium eccentricity by a prograde circumbinary disk \citep{Zrake+2021, MuditGarg:2023:LISA_e_Detect}}. 
Apsidal precession is then best understood as a change in orbital frequency of magnitude $\sim \pomega{b}$.
However, for a monochromatic source with an observed GW frequency $f_{\rm GW}$, there would be no simple way to distinguish between a vacuum source emitting at the ``Keplerian'' frequency $f_{\GW} / 2 = f_k$, or at a slightly perturbed frequency 
\begin{align}
     \frac{f_{\GW}}{2} = f_p = f_k \left( 1 + \frac{\Delta f}{f_k}\right),
 \label{eq:precess-freq}
\end{align}
where $\Delta f = (2\pi P_k)^{-1} \left<\delta \theta\right>_P \approx q_D f_k$, assuming the binary eccentricity is small and $\pomega{sim} \sim 1$. 
The two sources would generally only be distinguishable by their frequency evolution (or chirping) because they radiate gravitational energy at slightly different rates. Here we provide a simple estimate of this effect.

The radiated energy flux is $\dot E_\GW \sim f_\GW^6 a^4 \sim  \dot f$ such that a binary observed at $f_\GW$ that is perturbed by surrounding gas will exist at a slightly modified separation $a_p(f_\GW) \sim (2\pi f_p)^{-2/3}$ (compared to that on a Keplerian orbit in vacuum, $a_k$).
Thus, the modified chirp evolution can be expressed
\begin{align}
     \frac{\dot{f}_p}{\dot{f}_k} \sim \frac{a_p (f_{\rm GW})^4}{a_k(f_{\rm GW})^4} \sim \left(1 + q_D\right)^{- 8/3} \ .
\end{align}
The total accumulated phase of a GW signal is $\phi(f_\GW) = 2\pi \int f / \dot f_p df$
such that the difference in accumulated phase between the perturbed source and one in vacuum (for small $\Delta f$) is
\begin{align}
    \delta\phi = \phi_\GW - \phi_k \approx \frac{16\pi}{3} q_D \phi_k
 \label{eq:dephasing-approx}
\end{align}
where $\phi_k \sim f^{-5/3}$ is the total phase accumulated by a source in vacuum.
Such a dephasing is considered detectable in the LISA band (without accounting for degeneracies) if $\delta\phi \gtrsim 10 \times \SNR^{-1}$ with $\SNR$ the signal-to-noise ratio of the event \citep{Kocsis2011:EMRIs}. 
Thus, Equation~\ref{eq:dephasing-approx} gives a detectability condition for the disk mass as a function of the event $\SNR$ and the total number of orbits in-band $N \sim \phi_k / 2\pi$,
\begin{align}
    q_D \gtrsim \frac{3}{32\pi^2} \frac{10}{\SNR} N^{-1} \approx \frac{0.1}{N \times SNR} \ .
 \label{eq:detectable-gw-dephasing}
\end{align}
Equal mass mergers in the LISA band will have typical SNRs of order $\mathcal{O}(10^1-10^3)$ and will spend $N \sim 10 - 10^3$ orbits in band (with the exception of the most massive $10^7 \Msun$ binaries which will have $\SNR \sim 10 - 100$ and $N \sim 1 - 10$).

\begin{figure}[t!]
    \includegraphics{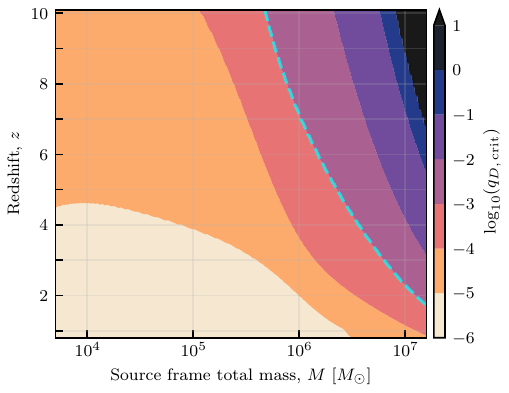}
    \vspace{-10pt}
    \caption{The critical disk mass $q_{D,\,\rm crit}$ required to induce a detectable dephasing in a LISA source with source frame total mass $M$ at redshift $z$.
    The \emph{blue-dashed} line indicates the viscous limit $q_D \sim 10^{-3}$ determined in Section \ref{s:viscous-limit}.
    }
 \label{fig:detectable-gw-dephasing}
\end{figure}

Figure~\ref{fig:detectable-gw-dephasing} illustrates the critical disk mass $q_{D, {\rm crit}}$ necessary to satisfy Equation~\ref{eq:detectable-gw-dephasing} as a function of the total source frame binary mass $M$ and redshift $z$ by calculating $\SNR(M, z)$ and $\phi_k(M, z)$ from PhenomA model waveforms \citep{Ajith:2007:PhenomA} and an approximate LISA sensitivity curve \citep{Robson:2019:LISA-SC}.
The \emph{blue-dashed} line indicates the $q_D = 10^{-3}$ viscous limit determined from Section \ref{s:viscous-limit} such that the majority of LISA sources are detectable for $q_D \gtrsim 10^{-5}$; and the most nearby sources for $q_D \gtrsim 10^{-6}$.
However, this estimate serves only as a lower bound on the disk mass required to induce a detectable dephasing in LISA\footnote{However, LISA progenitors are canonically thought to possess CBDs an order of magnitude or two thinner than those considered for this study, and such CBDs in steady-state have been associated with larger density pile-ups at the cavity edge \citep[e.g.][]{Rafikov:2016, Tiede:2020, Dittmann:2022:MachStudy}. 
The resultant increased outward radial force (see Section~\ref{S:Results}) could amplify the precession rate and the detectability of such observational effects.}
because we have ignored degeneracies in the induced chirp behavior {\color{magenta} (e.g. \citealt{MuditGarg:2022:LISA_dephasing})}, have assumed each source spends the maximum amount of time in-band possible for a 4 year mission lifetime, 
and have neglected the disk mass evolution with shrinking semi-major axis;
and for the lowest values of $q_D$ this signal may start to compete with the frequency resolution of the detector.
Furthermore, a more rigorous treatment ought to self-consistently connect the circumbinary disk forces to the time evolution of the binary quadrupole moment and the resultant chirp behavior.
We reserve this calculation for future work, but emphasize that because disk-induced precession dominates over the other orbital perturbations, it is the most likely candidate for detectable disk-induced signals in LISA detections of near equal-mass binaries.

\section{Conclusions} \label{S:Conclusions}

We have calculated from viscous hydrodynamics simulations the apsidal precession of eccentric binaries accreting from circumbinary disks.
Such simulations have previously been used to compute the disk-driven rate of the change to the binary semi-major axis and eccentricity, and we have found that the induced precession effect is an order of magnitude faster than these changes (Figure~\ref{fig:orb-evo}).
Moreover, we identify the primary source of this precession and the fundamental reason for its comparative significance as the relatively strong and symmetric outward radial gravitational force from the bulk of the circumbinary disk (Figure~\ref{fig:radial-forces} \& \ref{fig:integrated-forces}).

The degree to which $\pomega{b}$ competes with other timescales in the problem scales with the disk-to-binary mass ratio $q_D$, and for sufficiently large disk masses it is possible this precession could alter the full solutions on which previous orbital-evolution results and our own findings are based.
We estimated these disk masses by first comparing $\pomega{b}$ to the precession of a test particle in the quadrupole moment of the binary's potential and of the circumbinary disk itself (Figure~\ref{fig:binary-precession}).
We found that the disk-induced binary precession generally dominates over these timescales when $q_D \gtrsim 1 / 300$ with the exception of binaries with $0.2 \lesssim e_b \lesssim 0.4$ where the circumbinary disk is empirically found not to precess.
We compared the timescale for $\pomega{b}$ to the viscous time in the disk (Figure~\ref{fig:qd-visc}) and found that for disk masses $q_D \gtrsim 10^{-3}$ the induced binary precession may occur faster than the disk can viscously relax to the changing binary apsides, and that such situations may warrant more dedicated study to determine if the precession of the binary alters the full solution.
We additionally computed physical values of $q_D$ from $\alpha$ and $\beta$ steady-state disk solutions (Figure~\ref{fig:ab_disk_qd}) and determined that circumbinary accretion solutions and orbital evolution modelling for binaries with masses up-to $\sim 10^7 M_\odot$ and at sub-parsec separations may need to consider the effects of disk-induced binary precession.

For observational purposes, we compared the disk-induced binary precession to precession rates from general relativity (Figure~\ref{fig:binary-v-gr}) and discussed implications for both electromagnetic and gravitational wave searches for accreting and coalescing massive binaries.
Specifically, we determined that disk-induced precession may be significant over only a few orbital periods for modelling sources with Doppler-lensing periodic modulations, and that the precession may source an extra phase accumulation in accreting LISA systems that is generally detectable (in the best case scenario and ignoring degeneracies) for equal mass binaries with disk-binary mass ratios $q_D \gtrsim 10^{-5}$.
Thus, we have concluded that the effects of precession on existing solutions for binary orbital evolution and circumbinary accretion signatures warrants future, more detailed investigation, and that it should be considered in both current and future observations of accreting massive binaries in electromagnetic surveys and  gravitational wave experiments.

%
\section{Acknowledgments}
The authors sincerely thank Bin Liu, Alessandro Trani, Andrew MacFadyen, and members of the NBIA theoretical astrophysics group for insightful discussions.
D.J.D. and C.T. received support from the Danish Independent Research Fund through Sapere Aude Starting grant No. 121587.  
L.Z. acknowledges support from  ERC Starting grant No. 121817–BlackHoleMergs. 
P.D. acknowledges support from the National Science Foundation under grant AAG-2206299.

\bibliographystyle{mnras}
\bibliography{refs}

%
\appendix
\renewcommand\theequation{A\arabic{equation}}
\renewcommand\thefigure{A\arabic{figure}}
\setcounter{equation}{0}
\setcounter{figure}{0}
%
\section{$\pomega{b}$ derivation with accretion} \label{A:derivation}

Starting with the elliptic equation
\begin{align}
    r = \frac{a(1 - e_b^2)}{1 + e_b \cos{\nu}}
\end{align}
with $\nu = \theta - \varpi_b$ the true anomaly, $\theta$ the position angle measured from the line of nodes, $\varpi_b$ the longitude of pericenter, $a$ the semi-major axis, and $e_b = \sqrt{1 + 2\ell^2 E \mu^{-2}}$ the orbital eccentricity. $\ell = \sqrt{\mu a(1 - e_b^2)}$ is the specific angular momentum, $E = -\mu / 2a$ is the specific orbital energy, and $\mu = GM$ with $M$ the total mass.
Therefore, we write
\begin{align}
 \begin{split}
     \ell^2 &= \mu r (1 + e_b \cos{\nu}) \\ 
           &= \mu r \left[ 1 + \sqrt{1 + 2\ell^2 E \mu^{-2}} \cos{(\theta - \varpi_b)} \right] \ .
 \end{split}
\end{align}
Differentiating with respect to time while keeping the instantaneous radius fixed (e.g. because of some external force) yields
\begin{align}
 \begin{split}
    2\ell \dot \ell = \dot \mu r (1 &+ e_b \cos{\nu}) + \mu r e_b^{-1} \left[ \frac{E}{\mu^{2}}2\ell \dot\ell + \frac{\ell^2}{\mu^{2}}\dot E - \frac{2\ell^2 E}{\mu^{2}} \frac{\dot \mu }{\mu}  \right] \cos{\nu}  
    + \mu r e_b \left[  - \dot \theta + \dot \varpi_b \right] \sin{\nu} \ .
 \label{eq:derivative}
 \end{split}
\end{align}
The energy and angular momentum derivatives are given as
\begin{align}
 \begin{split}
     &\dot E = \frac{\mu}{2a^2}\dot a - \frac{\dot \mu}{2a} \\
    2 \ell &\dot \ell = \dot \mu a (1 - e_b^2) + \mu \dot a (1 - e_b^2) - 2\mu a \edot_b
 \end{split}
\end{align}
such that all terms proportional to $\dot \mu$ in Equation~\ref{eq:derivative} cancel. 
Therefore, as expected, there is no contribution to the binary precession from the accretion of mass.
Solving for $\pomega{b}$, then, gives
\begin{align}
    \nonumber
    \pomega{b} &= \dot \theta + \frac{2\ell \dot{\tilde\ell}}{\mu r e_b \sin{\nu}} - \frac{\cot{\nu}}{e_b^2 \mu^{2}} \left[ 2E\ell \dot{\tilde{\ell}} + \ell^2 \dot{\tilde E} \right] \\
    &= \dot \theta + \left( \frac{r^{-1} - E(e\mu)^{-1} \cos{\nu}}{\mu e_b \sin{\nu}} \right) 2\ell \dot{\tilde{\ell}} - \frac{\ell^2 \cot{\nu} }{\mu^2 e_b^2} \dot{\tilde E}  \ .
 \label{eq:solve-for-pomdot}
\end{align}
where $\dot{\tilde \ell}$ and $\dot{\tilde E}$ are the change in energy and specific angular momentum due only to an external specific force $\mathbf{f}$
\begin{align}
 \begin{split}
     \dot{\tilde E} &=  f_r v_r + f_\phi v_\phi = \sqrt{\frac{\mu}{a (1 - e_b^2)}} \bigg[  f_r e_b \sin{\nu} + f_\phi (1 + e_b \cos{\nu} ) \bigg]  \\
    \dot{\tilde \ell} &= r f_\phi 
 \end{split}
\end{align}
where we've used the fact that $v_r = \sqrt{\mu a^{-1} /(1 - e_b^2)} e_b \sin\nu$ and $ v_\phi = \sqrt{\mu a^{-1} / (1 - e_b^2)} (1 + e_b\cos\nu)$.
Lastly, we take $\dot \theta = 0$ because we are only considering in-plane forces, fixing the binary longitude of ascending node \citep{SolarSystemDynamics}.
Plugging in to Equation~\ref{eq:solve-for-pomdot} yields
\begin{align}
    \dot \varpi_b = e_b^{-1} \sqrt{a\mu^{-1} (1 - e_b^2)} \left[ -f_r \cos{\nu} + f_\phi \sin{\nu} \frac{2 + e\cos{\nu}}{1 + e\cos{\nu}} \right] \ .
\end{align}
%

\section{Retrograde CBDs} \label{A:retro}

\cite{TiedeDorazio:2023} also explored the orbital response of equal-mass binaries accreting from retrograde CBDs.
For completeness, we measure the disk-induced precession for retrograde solutions; this is shown in Figure~\ref{fig:prog-retro} as the \emph{blue} curve alongside $\pomega{b}$ from the prograde configuration.
Despite the comparative symmetry and lack of resonances in the retrograde scenario (see \citealt{TiedeDorazio:2023} for a detailed discussion), a retrograde CBD still drives \emph{prograde} apsidal precession in the binary.
This disk-induced precession in retrograde configurations is $\sim 2\times$ faster than in prograde disks at low binary eccentricity $e_b \lesssim 0.15$; but decreases monotonically with growing $e_b$ such that the effect is comparable by $e_b \gtrsim 0.4$.
The comparatively large precession rate in retrograde solutions is consistent with the relatively large magnitudes measured for $\dot a / a$ and $\edot_b$, and we posit that these effects are due to stronger gravitational forcing associated with a less truncated CBD (especially at low $e_b$) that is more tightly coupled to the binary. 
Nevertheless, all analysis presented in this paper would change by at most a factor of two if considering retrograde instead of prograde accretion scenarios.

\section{$\pomega{b}$ fitting functions} 
\label{A:fit}

\begin{figure}
    \vspace{12pt}
    \centering
    \includegraphics{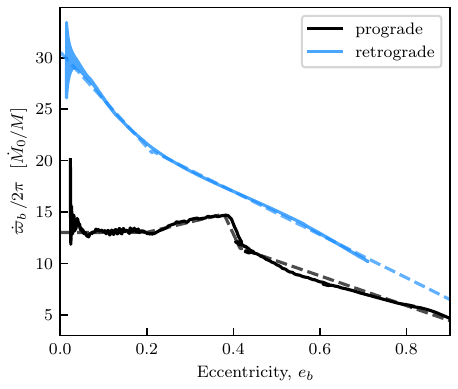}
    \caption{Binary precession from a retrograde CBD compared to the prograde solution.
    Fitting functions provided in Appendix C are shown by the associated \emph{dashed} lines. 
    }
 \label{fig:prog-retro}
\end{figure}

Figure~\ref{fig:prog-retro} also includes as \emph{dashed} lines approximate fitting functions for $\pomega{b}$ in both prograde (\emph{black}) and retrograde (\emph{blue}) configurations.
The piecewise fitting function for prograde solutions is taken as
\vspace{2pt}
\begin{eqnarray}
\pomega{fit} =   \frac{\dot M_0}{M}
  \begin{cases} 
      B_0 & e < e_0 \\
      B_0 + m_1 (e-e_0) & e_0 \leq e < e_1 \\
      B_1 + m_2 (e-e_1) & e_1 \leq e < e_2 \\
      B_2 + m_3 (e-e_2) & e > e_2 
   \end{cases}
\end{eqnarray} 
\vspace{2pt}
where $B_0=0.770$, $B_1 = 0.878, B_2 = 0.680$, 
$m_1 = 0.600$, $m_2 = -5.351, m_3 = -0.860$, and $e_0 = 0.198$, $e_1 = 0.378$, $e_2 = 0.415$.
For retrograde solutions we approximate the disk-induced precession as
\vspace{2pt}
\begin{eqnarray}
\pomega{fit\_r} =   \frac{\dot M_0}{M}
  \begin{cases} 
      C_0 + n_0 e & e < e_0^\prime \\
      C_1 + n_1 e & e \geq e_0^\prime
   \end{cases}
\end{eqnarray} 
\vspace{2pt}
where $C_0=1.810$, $C_1 = 1.503$, 
$n_0 = -2.738$, $n_1 = -1.242$, and $e_0^\prime = 0.212$.
%

\section{Phase Accumulation from Osculating Orbits}
\label{A:dphi}

The angle tracking the true position of the orbit in the sky is $\theta = \nu + \varpi_b$, where the true anomaly $\nu$ denotes the angle with respect to the argument of pericenter $\pomega{b}$. 
The true anomaly under perturbations can be expressed,
\begin{equation}
    \dot{\nu} = \dot{\nu}_k - \pomega{b},
\end{equation}
where $\dot{\nu}_k$ is the Keplerian value of the true anomaly advance.
However, by definition 
\begin{equation}
\dot{\theta} = \dot{\nu} + \pomega{b},
\end{equation}
so that $\dot{\theta} = \dot{\nu}_k$.
This implies that the only variation in $\dot{\theta}$ is caused by the the perturbed Kepler equation. 
Thus, under perturbed orbital elements $\chi_p = \chi_0 + \chi_1$, the rate at which the binary sweeps out this true longitude can be expressed
\begin{equation}
    \dot{\theta} = \dot\nu_k(\chi_p) = \sqrt{\frac{\mu}{p^3_p}}\left( 1 + e_p \cos{\nu_p}\right)^2,
 \label{eq:thetadot}
\end{equation}
where $p=a(1-e^2)$, and $\chi_1 = \int^t_{t_0}\dot{\chi}dt$ is proportional to the magnitude of the perturbing force and so assumed small over an orbit.
We expand the true anomaly by writing $\dot{\nu} = \dot{\nu_{k0}} - \pomega{b}$ and integrate---without yet specifying the time dependence of $\pomega{b}$---so that $\nu_p(t) = \nu_{k0}(t) - \varpi_b(t)$. Here $\nu_{k0}$ is shorthand for $\nu_k(\chi_0)$. 
We expand in the limit of slow precession and small perturbations, keeping only quantities to first order in these small quantities (e.g. $a_p = a_0 + a_1$ and so forth) such that Equation~\ref{eq:thetadot} becomes
\begin{eqnarray}
    \dot{\theta} = \dot \nu_{k0} + \dot \nu_{k0}\left[ \left( -\frac{3}{2}\frac{a_1}{a_0} + \frac{3e_0e_1}{1-e^2_0} \right) + \frac{2}{1+e_0\cos{\nu_{k0}}} \left( e_1\cos{\nu_{k0}} + e_0 \varpi_b(t) \sin{\nu_{k0}} \right) \right]. 
\end{eqnarray}
Dephasing is from the second term, and $\varpi_b(t)$ enters only in the last term.
Expanding the phase-dependent $\cos$ and $\sin$ terms in time via the mean anomaly $\tilde M = \Omega_b t$ to 1st order in $e_0$, we write
\begin{equation}
\delta \dot{\theta}(t) = 2 \sqrt{\frac{\mu}{p^3_0}}\left[  e_0 \varpi_b(t) \sin{\tilde M}  - \frac{3}{4}\frac{a_1}{a_0}\left( 1 + 2e_0 \cos{\tilde M}\right) 
+ e_0 e_1 \left( 1 + \frac{3}{2} \cos{2 \tilde M}\right)  \right].
\end{equation}

Now consider the contribution from each of the perturbed quantities after one orbit, $\varpi_b(P_b)$, $a_1(P_b)$, and $e_1(P_b)$, in the small $e_0$ limit where orbital dephasing is most relevant for LISA sources.
To see the contribution from $\varpi_b(P_b)$ we expand $e_0 \pomega{b}$ as,
\begin{align}
      e_0 \pomega{b} &\approx \sqrt{\frac{p_0}{\mu}} \left[ -f_r \cos{\nu_{k0}} + 2f_{\phi} \sin{\nu_{k0}} \left(2 - e_0 \cos{\nu_{k0}} \right) \right] + \mathcal{O}(e^2_0) \nonumber\\
      %
      %
      &\approx \sqrt{\frac{p_0}{\mu}} \left[ -f_r \cos{\tilde M} +  4 f_{\phi}\sin{\tilde M} \right] + \mathcal{O}(e_0)
\end{align}
where we've once again expanded the oscillating functions of true anomaly in mean anomaly (time) to linear order in $e_0$.
Then the leading order contribution to $\varpi_b$ and the dephasing will depend on the functional dependence of the forces. 

For prograde, coplanar accretion the radial force dominates $f_r \gg f_\phi$ and for small $e_0$ is almost constant in time.
For larger $e_0$, $f_r$ has significant $\cos{\tilde M}$ components (see Figure~\ref{fig:radial-forces}).
Therefore, we consider two specific scenarios: that of constant radial force $f_r=A_r$ and that where $f_r = -A_r \cos{\tilde M}$.
For each of these scenarios $a_1/a_0, \, e_0 e_1 \sim f_\phi + \mathcal{O}(e_0)$ (as discussed in \S \ref{S:Results}), and the dominant term in $\delta \dot \theta$ is the precession term.

For the constant force, the largest contributions to $\varpi_b$ will be those that do not vanish in the integral over an orbit. To leading order,
\begin{eqnarray}
    e_0\varpi_b(t) =  -\frac{a^2_0}{\mu} A_r \sin{\tilde M}  + \mathcal{O}(e_0, f_{\phi})
\end{eqnarray}
and the dephasing over an orbit is given by
\begin{eqnarray}
\delta {\theta}(P_b) &\approx& 2 \sqrt{\frac{\mu}{p^3_0}} \int^{P_b}_0  e_0 \varpi_b(t) \sin{\tilde M}  dt \nonumber \\
%
%
&\approx&  - 2 \pi \frac{A_r a^2_0}{\mu}  \propto  -2 \pi q_D,
\end{eqnarray}
where we've used $A_r / (\mu / a_0^2) \sim q_D$ because the force is proportional to the disk mass.

For, instead, $f_r = -A_r \cos{\tilde M}$,
\begin{align}
    e_0\varpi_b(t) =  \frac{1}{4}\frac{a^2_0}{\mu} \left[ A_r \left(2 \tilde M + \sin{2 \tilde M}\right) \right]  + \mathcal{O}(e_0, f_{\phi})
\end{align}
and 
\begin{eqnarray}
    \delta {\theta}(P_b) \approx  -2 \pi \frac{A_r a^2_0}{\mu} \propto -2 \pi q_D.
\end{eqnarray}
Hence, for both scenarios, the dephasing from apsidal precession is to leading order given by $2\pi q_D \sim \pomega{b} P_b$.

\end{document}